\documentclass[12pt]{iopart}
\usepackage{iopams}
\usepackage{graphicx,color}

\newcommand{\be}{\begin{equation}}
\newcommand{\ee}{\end{equation}}
\newcommand{\bea}{\begin{eqnarray}}
\newcommand{\eea}{\end{eqnarray}}
\newcommand{\kref}[1]{(\ref{#1})}

\newcommand{\ds}{{\sf DarkSUSY}}

\begin{document}

\title[Particle DM and small-scale structure]{
Particle Models and the Small-Scale Structure\\ of Dark Matter}

\author{Torsten Bringmann}

\address{
Oskar Klein Centre for Cosmo Particle Physics, Department of Physics, Stockholm University, AlbaNova, SE - 106 91 Stockholm, Sweden
}
\ead{troms@physto.se}

%Mearth/Msun=3x10^-6

\begin{abstract}
The kinetic decoupling of weakly interacting massive particles (WIMPs) in the early universe sets a scale that can directly be translated into a small-scale cutoff in the spectrum of matter density fluctuations. The formalism presented here allows a  precise description of the decoupling process 
and thus the determination of this scale to a high accuracy from the details of the underlying WIMP microphysics.
With decoupling temperatures of several MeV to a few GeV, the smallest protohalos to be formed range between $10^{-11}$ and almost $10^{-3}$ solar masses -- a somewhat smaller range than what was found earlier using order-of-magnitude estimates for the decoupling temperature; for a given WIMP model, the actual cutoff mass is typically about a factor of 10 greater than derived in that way, though in some cases the difference may be as large as a factor of several 100. 
Observational consequences and prospects to probe this small-scale cutoff, which would provide a fascinating new window into the particle nature of dark matter, are discussed.

\end{abstract}

%\submitto{\NJP}

\maketitle

\section{Introduction}

The combined data of cosmic microwave observations \cite{wmap}, distance measurements of type Ia supernovae \cite{sn} and the baryon acoustic oscillations in the distribution of galaxies \cite{bao}, along with a wealth of other observations, provide ample evidence for a minimal, 6-parameter cosmological $\Lambda$CDM model, where the dominant contribution to the total energy density of the universe today are dark energy and cold dark matter (CDM). While the nature of the latter (and even more so of the former) remains unknown, weakly interacting massive particles (WIMPs) provide a class of particularly well motivated CDM candidates that, thermally produced in the early universe, naturally acquire the necessary relic density today (for reviews on particle DM, see \cite{JKG,reviews}). It is interesting to note that WIMPs automatically arise as a by-product of almost any attempt to address the problems of the standard model of particle physics (SM) that become apparent at the TeV scale, viz.~its extreme UV sensitivity in the scalar sector, a scale that soon will be accessible for the first time by the LHC. The most popular, and arguably most well-motivated scenario in this context is supersymmetry \cite{JKG}, but interesting alternatives include, e.g.,  scenarios with large extra dimensions \cite{ued} or little-Higgs models \cite{Birkedal:2006fz}. For concreteness, the focus will often be on the supersymmetric neutralino, but the main results -- such as the mass range of the smallest WIMP protohalos -- will be equally applicable to, and include, other WIMP DM candidates.

After the WIMP number density has frozen out in the early universe, setting the DM relic density today, scattering events with the much more abundant SM particles still keep the WIMPs extremely close to thermal equilibrium. The complete decoupling from the thermal bath happens only considerably later \cite{Schmid:1998mx,Chen:2001jz}, when the temperature has dropped by another factor of $\sim10-1000$, from which moment on the DM particles can stream freely and thereby wash out any matter density contrasts on small scales \cite{Hofmann:2001bi,Berezinsky:2003vn,Green:2003un,Green:2005fa}. Acoustic oscillations during and after kinetic decoupling also give rise to a damping of the power spectrum of matter density fluctuations \cite{Loeb:2005pm,Bertschinger:2006nq}; whether or not this damping is stronger than the one induced by free streaming, as we will see later, depends on the details of the WIMP model. In any case is it the kinetic decoupling scale that determines the cutoff in the density spectrum and thus provides a direct link between the size of the smallest gravitationally bound structures and the CDM particle nature. To make proper use of this link, it is necessary to develop a detailed understanding of the decoupling process so as to allow a precise determination of the decoupling temperature.

This article is organized as follows. Section \ref{sec:chi} briefly introduces the supersymmetric neutralino as DM candidate, which for most of the article will be used as a prototype for the more general class of WIMP DM.
In Section \ref{sec:dec}, a detailed description of the decoupling of WIMPs from the thermal bath is given, together with a prescription for a highly accurate determination of the kinetic decoupling temperature. The subsequent evolution of matter density fluctuations until the formation of the first protohalos is then discussed in Section \ref{sec:mcut}. Observational consequences and prospects to probe the power spectrum at the smallest scales are outlined in Section \ref{sec:disc}. Here, also the situation for other (WIMP and non-WIMP) DM candidates are briefly discussed. Section \ref{sec:con}, finally, concludes. The two appendices provide analytic expressions for the collision term in the Boltzmann equation, describing the scattering between arbitrary DM candidates and SM heat bath particles, as well as the full scattering amplitude for neutralino-fermion scattering in supersymmetry.

%%%%%%%%%%%%%%%%%%%%%%%%%%%%%%%%%%%%%%%%%%%%%%%%%%%%%%%%%%%%%%%%%%%%%%%%%%%%%%
\section{Supersymmetry and neutralino dark matter}
\label{sec:chi}

Major motivations to consider supersymmetric (SUSY) extensions to the SM include the (at least in the unbroken theory) perfect symmetry between fermions and bosons, which not only successfully addresses the unification of gauge couplings at the scale of grand unified theories (GUT) but also, for realistic schemes of SUSY breaking, the aforementioned hierarchy problem of the SM. The conservation of $R$-parity, introduced to ensure the stability of the proton against decay, means that the lightest SUSY particle (LSP) has to be stable. In most models, the LSP is the lightest neutralino, a linear combination of the superpartners of the gauge and Higgs fields,
\be
  \chi\equiv\tilde\chi^0_1= N_{11}\tilde B+N_{12}\tilde W^3 +N_{13}\tilde H_1^0+N_{14}\tilde H_2^0\,,
\ee
 which provides a perfect WIMP DM candidate (for a review, see \cite{JKG}). 

A popular way of assessing the range of possible phenomenologies for neutralino DM is to study a $7$ parameter version (at the electroweak scale) of the minimal supersymmetric extension of the SM (MSSM), where
 $\mu$ denotes the Higgsino mass parameter, $\tan\beta$ the ratio of vacuum expectation
values of the two Higgs doublets, $M_1$, $M_2$ and $M_3$ the gaugino
mass parameters, and $m_A$  the neutral pseudoscalar Higgs mass. Further parameters are $m_0$, $A_t$ and $A_b$, which are defined through a
simplifying ansatz to describe the soft trilinear couplings  ${\bf A}$ and the soft sfermion masses ${\bf M}$ (which in general
are $3\times3$ matrices in generation space): ${\bf M}_Q = {\bf M}_U = {\bf M}_D = {\bf M}_E = {\bf M}_L = m_0{\bf 1}$,
${\bf A}_U = {\rm diag}(0,0,A_t)$, ${\bf A}_D = {\rm diag}(0,0,A_b)$, ${\bf A}_E = {\bf 0}$. No CP-violating phases other than the CKM phase of the SM are allowed for;
as a natural further simplification, the GUT condition for the gauge
couplings, leading to $M_1={5\over 3}\tan^2\theta_wM_2\approx {1\over 2}M_2$
is used. 

A more restricted, but in some sense more natural, set of parameters defines
models of minimal supergravity (mSUGRA), where gravity is supposed to mediate SUSY breaking.
In these models, parameters are given at the GUT scale; the mass spectrum and couplings at the electroweak scale are then calculated by solving the renormalization group equations. The parameters of mSUGRA models are the universal scalar and gaugino masses, $m_0$ and $m_{1/2}$,  a common trilinear term $A_0$, $\tan \beta$, and the sign of $\mu$. From a cosmological point of view, the regions in parameter space that are particularly interesting
-- as they correspond to models giving the correct relic density -- are: The \emph{bulk region} at low $m_0$ and $m_{1/2}$; the \emph{funnel region}
$m_A\approx 2m_\chi$, where
annihilations in the early universe are enhanced by the presence of the
near-resonant pseudoscalar Higgs boson; the hyperbolic branch or \emph{focus point region} where $m_0 \gg m_{1/2}$; the \emph{stau coannihilation region}
where $m_{\chi}\approx m_{\tilde\tau}$; and finally the \emph{stop coannihilation region} (arising when $A_0 \ne 0$) where $m_\chi \approx m_{\tilde{t}}$.

To assess the range of decoupling scales for neutralino DM, an extensive scan over the MSSM and mSUGRA parameter space was performed (the same as described in \cite{IB_SUSY}), resulting in about $4\cdot10^5$ models that satisfy all current bounds from accelerator physics and at the same time feature a relic density within the $3\sigma$ bound of the combined WMAP, supernovae and baryonic acoustic oscillation data \cite{wmap}:
\be
  \label{wmap}
    0.103<\Omega_\chi h^2<0.123\,.
\ee
For the numerical calculations in this article, the \ds\ code was used
(see \cite{ds} for sign conventions and other details) which implements FeynHiggsFast \cite{feynhiggs} for the Higgs boson masses and decay widths, and  Isajet \cite{isajet} for the solution of the renormalization group equations (RGE) and for the mass spectra in mSUGRA models. The routines for the calculation of the decoupling scale as described in the next Section will be available with the next release of \ds\ \cite{dsupdate}.

%%%%%%%%%%%%%%%%%%%%%%%%%%%%%%%%%%%%%%%%%%%%%%%%%%%%%%%%%%%%%%%%%%%%%%%%%%%%%%
\section{Thermal decoupling of WIMPs}
\label{sec:dec}

At high temperatures in the early universe, $T\gtrsim m_\chi$, chemical equilibrium is maintained by the detailed balance between WIMPs annihilating into SM particles and the reverse process of SM particles creating WIMPs. In addition, local thermal equilibrium is guaranteed by highly frequent elastic scattering processes. Quantitatively, the evolution of the WIMP phase-space density $f$ is described by the Boltzmann equation; denoting with $p^\mu=(E,\mathbf{p})$ the comoving WIMP momenta and assuming a flat Friedmann-Robertson-Walker spacetime, it reads:
\be
  \label{boltz}
  E(\partial_t-H\,\mathbf{p}\cdot\nabla_\mathbf{p})\,f=C[f]\,,
\ee
where $H=\dot a/a$ is the Hubble parameter and $a$ the scale factor. The right-hand side is a source term, also known as the collision term, that describes any changes in $f$ that cannot simply be attributed to the geometry of the expanding spacetime. 

Dividing Eq.(\ref{boltz}) by $E$ and integrating it over $\mathbf{p}$, it becomes the familiar evolution equation for the WIMP number density $n$:
\be
  \label{boltzn}
  \partial_tn+3Hn=a^{-3}\partial_t\left(a^3n\right)=-\langle\sigma v\rangle\left(n^2-n_{\rm eq}^2\right)\,,
\ee
where $n_{\rm eq}$ denotes the number density in chemical equilibrium. Note that  only non-number conserving contributions to $C[f]$ survive the integration over $\mathbf{p}$, which is the reason for the appearance of the thermally averaged annihilation rate $\langle\sigma v\rangle$ on the right-hand side. Incidentally, this equation takes the same form, after replacing $\langle\sigma v\rangle$ with an \emph{effective} annihilation rate, even when coannihilations with particles close in mass to $m_\chi$ are taken into account \cite{Edsjo:1997bg}. At high temperatures, the right-hand side of Eq.~(\ref{boltzn}) dominates the Hubble expansion term and the WIMP number density is forced to follow its equilibrium solution, $n\simeq n_{\rm eq}$, which for $T\lesssim m_\chi$ is well approximated by a Maxwell-Boltzmann distribution. As $T$ drops further, however, the Boltzmann suppression of $n$ (and $n_{\rm eq}$) means that the WIMP annihilation (and creation) rate will eventually drop below the Hubble expansion rate; at \emph{chemical decoupling}, $T=T_{\rm cd}$, the comoving number density $a^3n$ starts to deviate considerably from its equilibrium value and finally freezes out and stays constant; for WIMPs, this happens around $T_{\rm cd}\sim m_\chi/25$.

Even at temperatures below $T_{\rm cd}$, however, the WIMPs are  kept in local thermal equilibrium by scattering processes with SM particles (for which the full collision term is provided in \ref{app:coll}). As pointed out in \cite{BH}, it is sufficient for the discussion of the kinetic decoupling process to consider the second moment of the Boltzmann equation (instead of the first moment like for chemical decoupling): multiplying Eq.(\ref{boltz}) by $\mathbf{p}^2/E$, integrating it over $\mathbf{p}$ and keeping only the leading order terms in $\mathbf{p}^2/m_\chi^2$ results in
\be
  \label{boltz2}
  \left(\partial_t+5H\right)T_\chi=2\,m_\chi\, c(T)\left(T-T_\chi\right)\,,
\ee
where $c(T)$ is given in Eq.~(\ref{cTdef}) and $T_\chi$ is defined by
\be
  \int\frac{d^3p}{(2\pi)^3}\mathbf{p}^2f(\mathbf{p})\equiv3\,m_\chi T_\chi n_\chi\,.
\ee
Note that this ``temperature parameter'' for the WIMPs does not require any assumptions about the form of $f(\mathbf{p})$, but simply provides a convenient means of characterizing the deviation from thermal equilibrium (for which $T_\chi=T$ holds).

\begin{figure}[t]
   \includegraphics[width=0.502\columnwidth]{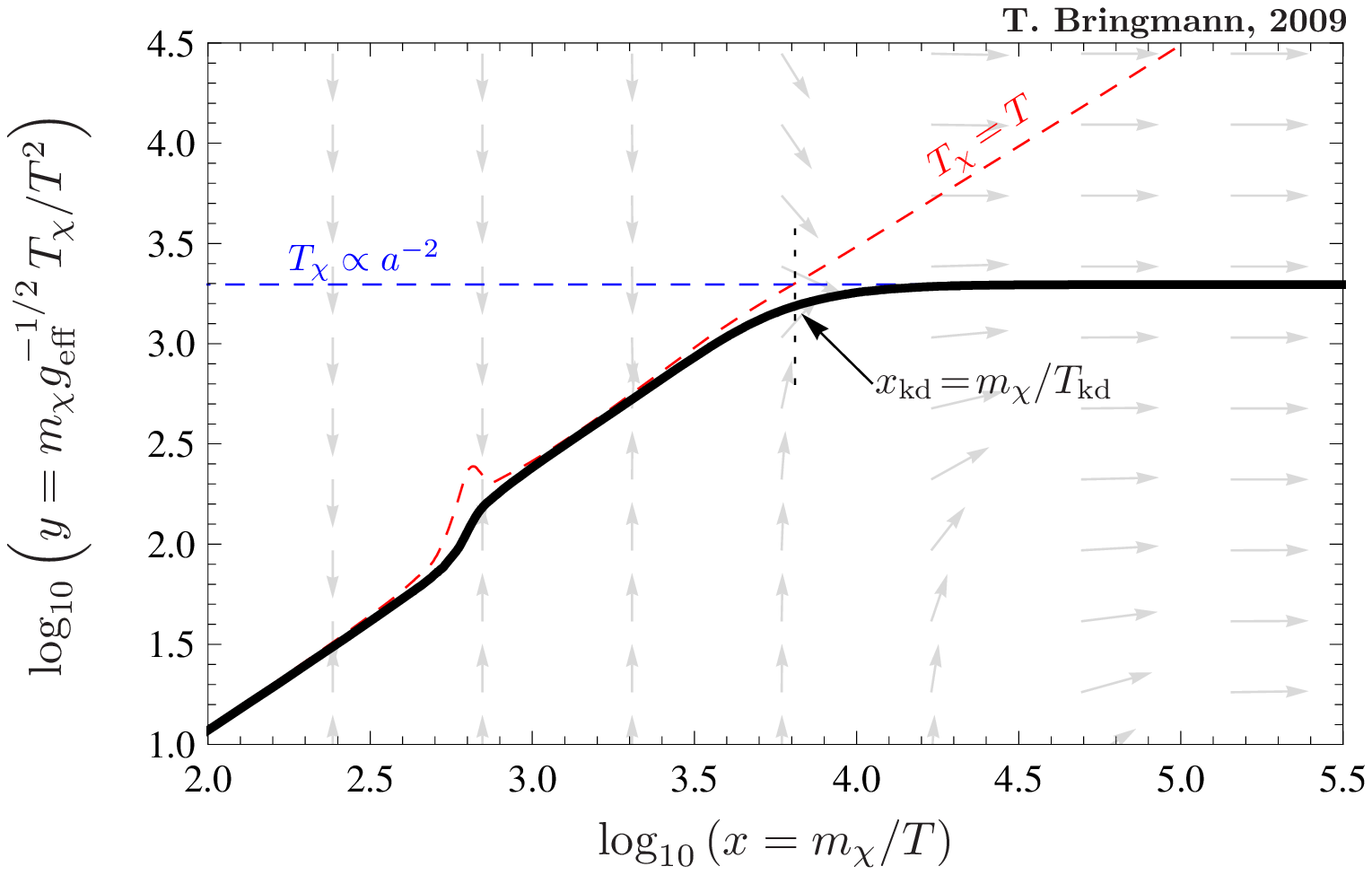}
   \includegraphics[width=0.468\columnwidth]{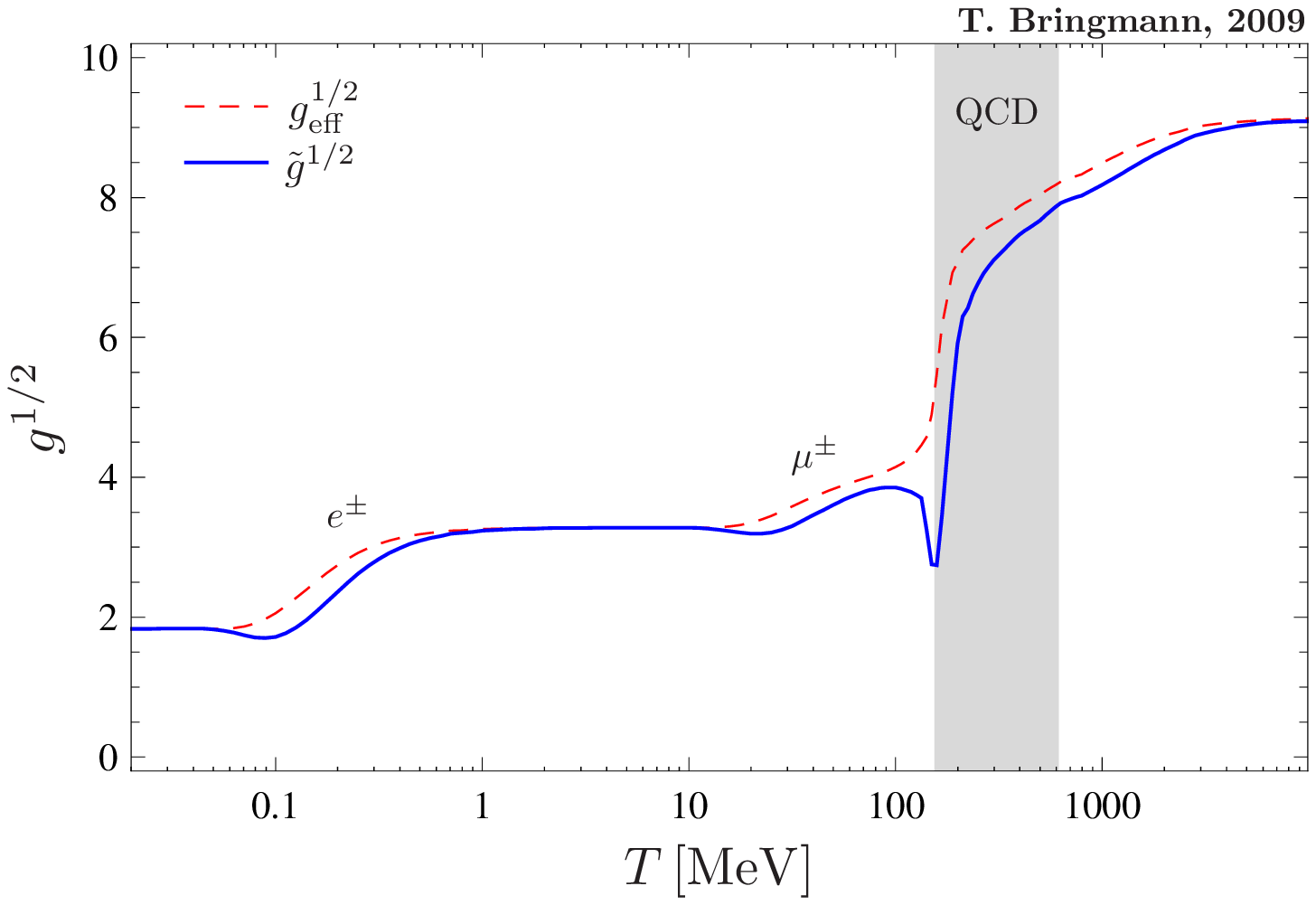}
\caption{The left panel shows the phaseplot and solution for the WIMP temperature evolution, for $m_\chi\sim100\,$GeV and $\overline{\left|\mathcal{M}\right|}^2\sim g_Y^4(m_\chi/\omega)^2$, expressed in the dimensionless variables introduced in Eqs.~(\ref{xdef},\,\ref{ydef}). At $T\lesssim T_{\rm kd}$, any departure from thermal equilibrium ($T_\chi=T$) is restored almost immediately (except for a short period around the QCD phase transition); for $T\gtrsim T_{\rm kd}$, the WIMPs decouple from the thermal bath and cool down with the Hubble expansion as $T_\chi\propto a^{-2}$.

In the right panel,  the effective number of relativistic degrees of freedom is plotted as a function of the temperature, implementing the results of \cite{Hindmarsh:2005ix} for the evolution of this quantity during the QCD phase transition; for reference, the decoupling of muons and electrons is also indicated.
 \label{fig_dof_phase}}
\end{figure}

During radiation domination, the expansion rate is given by
\be
  H^2=\frac{8\pi}{3m_{\rm Pl}^2}\rho_r=\frac{4\pi^3}{45m_{\rm Pl}^2}g_{\rm eff}T^4\,,
\ee
where $g_{\rm eff}$ is the effective number of relativistic degrees of freedom. Introducing further
\bea
   \label{xdef}
   x &\equiv& m_\chi/T\,,\\
   \label{ydef}
   y &\equiv& m_\chi g_{\rm eff}^{-1/2} T_\chi/T^2\,,
\eea
one can now bring Eq.~(\ref{boltz2}) into the convenient form
\be
  \label{eq:process}
  \frac{dy}{dx} = 2\frac{m_\chi\, c(T)}{H\tilde g^{-1/2}}\left(1-\frac{T_\chi}{T}\right)\,,
\ee
with $\tilde g^{1/2}\equiv g_{\rm eff}^{1/2}/(1+\frac{1}{4}\frac{g_{\rm eff}'}{g_{\rm eff}}T)$. From this equation, one can directly read off the asymptotic behaviour of the WIMP temperature: At large $T$, the term in front of the right-hand side is much larger than unity and one has $T_\chi=T$; when $T$ becomes small, the WIMPs finally decouple completely from the thermal bath and $y$ stays constant, i.e. $T_\chi\propto T^2 g_{\rm eff}^{1/2}\propto a^{-2}$. The \emph{kinetic decoupling} temperature is thus conveniently defined by equating these two regimes, as if the decoupling process were to occur instantaneously:
\be
  x_{\rm kd}=\frac{m_\chi}{T_{\rm kd}}\equiv g_{\rm eff}(T_{\rm kd})\, \left.y\right|_{x\rightarrow\infty}\,.
\ee
This procedure is illustrated in the left panel of Fig.~\ref{fig_dof_phase} which shows, for a typical WIMP candidate, the phaseplot and solution of Eq.~(\ref{eq:process}). As one can see, kinetic decoupling is indeed a process that happens on a rather short time scale; it becomes also apparent that thermal equilibrium is maintained extremely efficiently for $T\lesssim T_{\rm kd}$, with $T_\chi=T$ being an attractor solution that would be restored almost immediately after any departure from this behaviour (except for a short period during the QCD transition, see below, when the rapidly changing effective number of degrees of freedom does not allow this).

In principle, the scattering with all types of SM particles contributes to $c(T)$, see Eq.~(\ref{cTdef}). This picture is a bit complicated by the fact that kinetic decoupling in some cases can take place close to, or even above the QCD phase transition, the details of which are not yet fully understood. Lattice calculations, however, start to converge at a value for the critical temperature of $T_c\approx170\,$MeV for the most interesting case of two light (up and down) and one more massive (strange) quark flavour \cite{tqcd} and indicate that the plasma can be described by free quarks and gluons only for $T\gtrsim4T_c$ \cite{Boyanovsky:2006bf}. For the effective number of degrees of freedom during the transition, we adopt the results of \cite{Hindmarsh:2005ix} as displayed in the right panel of Fig.~\ref{fig_dof_phase}. As scattering partners are concerned, we conservatively restrict ourselves to leptons and, for $T>4T_c$, to the three lightest quarks.

\begin{figure}[t]
\begin{center}
   \includegraphics[width=0.483\columnwidth]{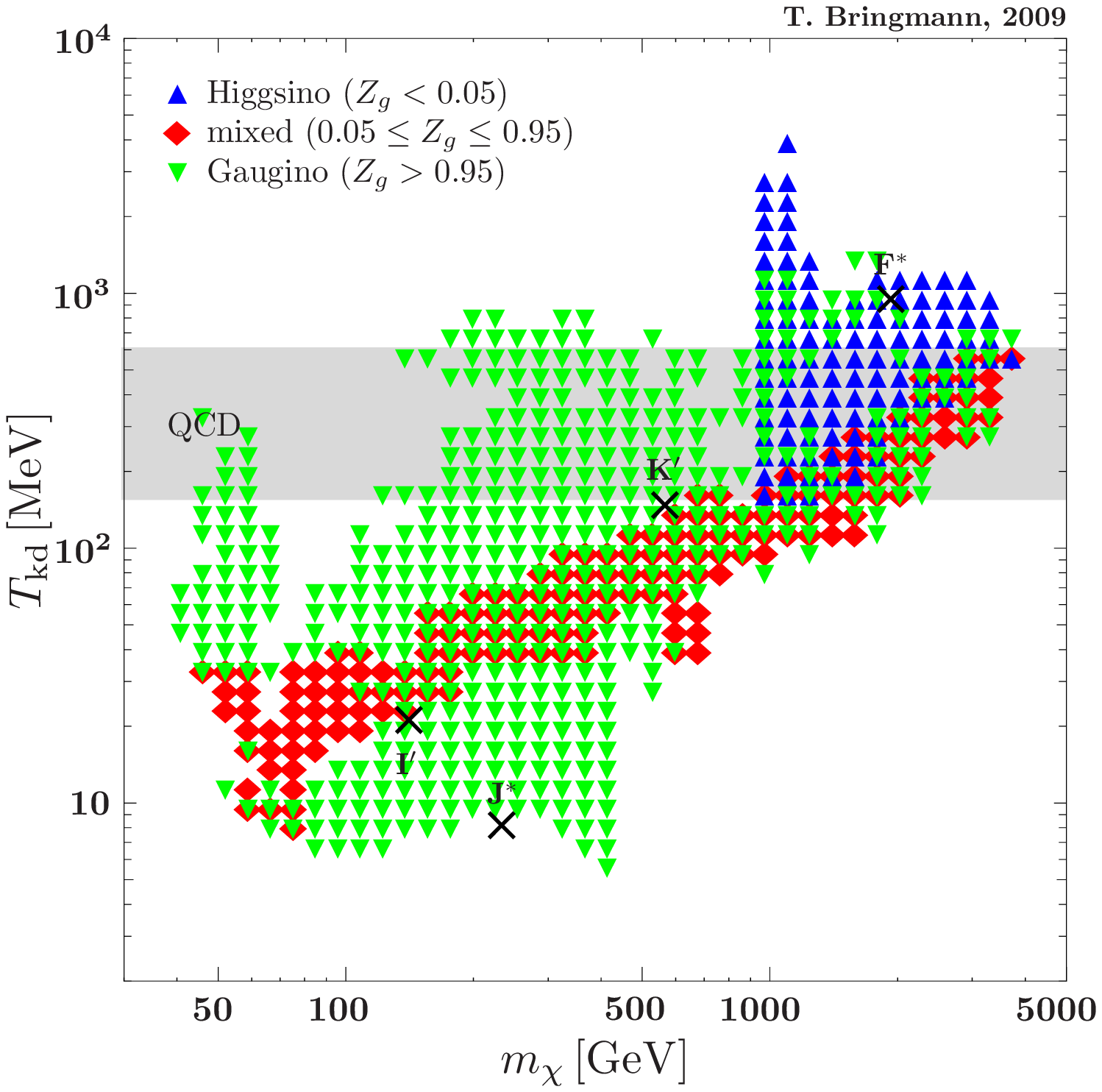}
   \includegraphics[width=0.477\columnwidth]{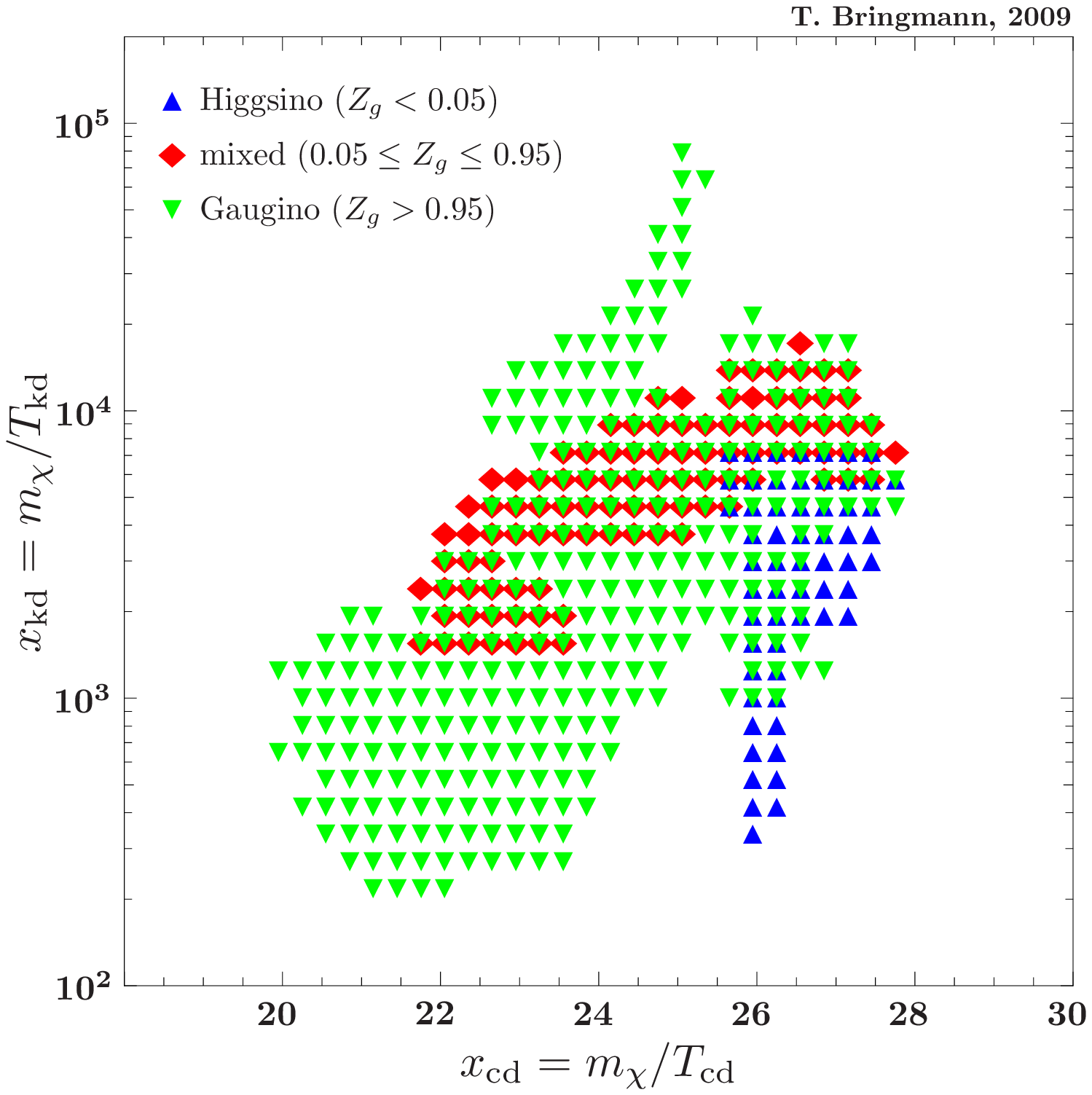}
\end{center}
\caption{The range of decoupling temperatures for neutralino DM. For models that fall inside or above the gray band marking the QCD phase transition, the actual value of $T_{\rm kd}$ will be slightly smaller than indicated. See text for further details.
\label{fig_Tkd}}
\end{figure}

The resulting range in $T_{\rm kd}$ for neutralino dark matter, obtained after having performed the extensive scan described in Section \ref{sec:chi}, is shown in Fig.~\ref{fig_Tkd} as a function of the mass $m_\chi$ and gaugino fraction $Z_g\equiv \left|N_{11}\right|^2+ \left|N_{12}\right|^2$ (in our case dominated by the Bino fraction). The gray band indicates the QCD phase transition; values for $T_{\rm kd}$ inside or above this band should be interpreted as \emph{upper bounds} on the decoupling temperature since the scattering with some of the hadronic degrees of freedom was not taken into account. On the other hand, as the coupling of WIMPs to hadrons is usually smaller than to leptons, the difference between this upper bound and the actual value of $T_{\rm kd}$ is not expected to be very big; note also that the scattering with bound QCD states like, e.g., pions is suppressed due to their rather large masses and thus small abundance (the evolution of density fluctuations, on the other hand, may very well be influenced by the details of the QCD phase transition, see the next Section). In addition to the result of the scan, the Figure also indicates the decoupling temperature for four mSUGRA benchmark models that were introduced in \cite{Battaglia:2003ab,IB_SUSY} and present typical examples for neutralinos in the bulk ($I'$), coannihilation ($J^*$), funnel($K'$) and focus point ($F^*$) region; the quite different annihilation spectra in gamma rays for these models, and the resulting prospects for indirect detection, were recently studied in some detail in \cite{IB_SUSY,IBobs}.

Assuming a constant equation of state and relativistic scattering partners, Eq.~(\ref{boltz2}) has actually an analytic solution \cite{BH} that can be used for a quick estimate of $T_{\rm kd}$. This estimate proves to be rather good (within 10\% of the full result shown here) for masses below a few hundred GeV; above that, however, the exact mass dependence of the number density of $\tau$ leptons, in particular, can be crucial, leading in some cases to differences of more than a factor of 5 between the two results.
A further observation is that MSSM and mSUGRA models occupy roughly the same regions in the $T_{\rm kd}$ -- $m_\chi$ plane; the largest values of $T_{\rm kd}$ for Higgsinos (at high masses) and Binos (at intermediate masses), however, corresponds nearly exclusively to MSSM models, while almost all Binos with a decoupling temperature below the band occupied by mixed neutralino are mSUGRA models.

For illustrative reasons, finally, the right panel of Fig.~\ref{fig_Tkd} compares the kinetic with the chemical decoupling temperature: as anticipated, kinetic decoupling takes place much later than chemical decoupling, at temperatures a factor of $10-1000$ lower. The possible range in $T_{\rm kd}$ is, furthermore, considerably larger than the one in $T_{\rm cd}$ -- which of course simply reflects the fact that the DM relic density is constrained extremely well while there are so far no observations that would put stringent bounds on $T_{\rm kd}$ (see also Section \ref{sec:obs}). Note also that $T_{\rm kd}$ is fairly uncorrelated with the chemical decoupling temperature, as well as with the neutralino annihilation cross section.

To conclude this Section, let us recall that the formalism presented here keeps the leading order terms in $\mathbf{p}^2/m_\chi^2$, thus allowing the determination of the decoupling scale to an accuracy of $\mathcal{O}\left(x_{\rm kd}^{-1}\right)$; while this is usually more than sufficient, it would be straightforward to include also higher orders in the discussion of the Boltzmann equation and thereby arrive at an even more accurate description of the decoupling process.

%%%%%%%%%%%%%%%%%%%%%%%%%%%%%%%%%%%%%%%%%%%%%%%%%%%%%%%%%%%%%%%%%%%%%%%%%%%%%%
\section{From decoupling to the first protohalos}
\label{sec:mcut}

Before kinetic decoupling, small-scale perturbations in the CDM fluid are damped due to the tight coupling of the WIMPs to the heat bath. For temperatures $T\lesssim T_{\rm kd}$, first a remaining viscous coupling between these two fluids and then the free-streaming of the WIMPs generate an exponential cutoff in the power spectrum \cite{Green:2005fa}, with a characteristic comoving damping scale of
\be
  k_{\rm fs}\approx\left(\frac{m_\chi}{T_{\rm kd}}\right)^{1/2}\frac{a_{\rm eq}/a_{\rm kd}}{\ln(4a_{\rm eq}/a_{\rm kd})}\frac{a_{\rm eq}}{a_0}H_{\rm eq}\,.
\ee
The WIMP mass contained in a sphere of radius $\pi/k_{\rm fs}$ is thus given by
\be
  \fl M_{\rm fs}\approx\frac{4\pi}{3}\rho_\chi\left(\frac{\pi}{k_{\rm fs}}\right)^3
=2.9\times 10^{-6}\left(\frac{1+{\rm ln}\left(g_{\rm eff}^{1/4}T_{\rm kd}/50\;{\rm MeV}\right)/19.1}{\left(m_\chi/100\; {\rm GeV}\right)^{1/2} g_{\rm eff}^{1/4}\left(T_{\rm kd}/50\;{\rm MeV}\right)^{1/2}}\right)^3M_\odot\,.
\ee
Later, it was shown that acoustic oscillations also have to be taken into account \cite{Loeb:2005pm,Bertschinger:2006nq}. Leading to a similar exponential cutoff, the characteristic damping mass in this case is given by the CDM mass inside the horizon at the time of kinetic decoupling:
\be
  M_{\rm ao}\approx\frac{4\pi}{3}\left.\frac{\rho_\chi}{H^3}\right|_{T=T_{\rm{kd}}}
  =3.4\times10^{-6}\left(\frac{T_{\rm kd}g_{\rm eff}^{1/4}}{50\,{\rm MeV}}\right)^{-3}M_\odot\,.
\ee
Note that the above expressions contain the full dependence on $g_{\rm eff}$ (evaluated at $T_{\rm kd})$, which is often not quoted despite the large range of possible decoupling temperatures; to arrive at the numerical values,
the most recent estimates for $\Omega_\chi h^2$ and $\Omega_m h^2$ \cite{wmap} were used.

\begin{figure}[t]
   \includegraphics[width=0.474\columnwidth]{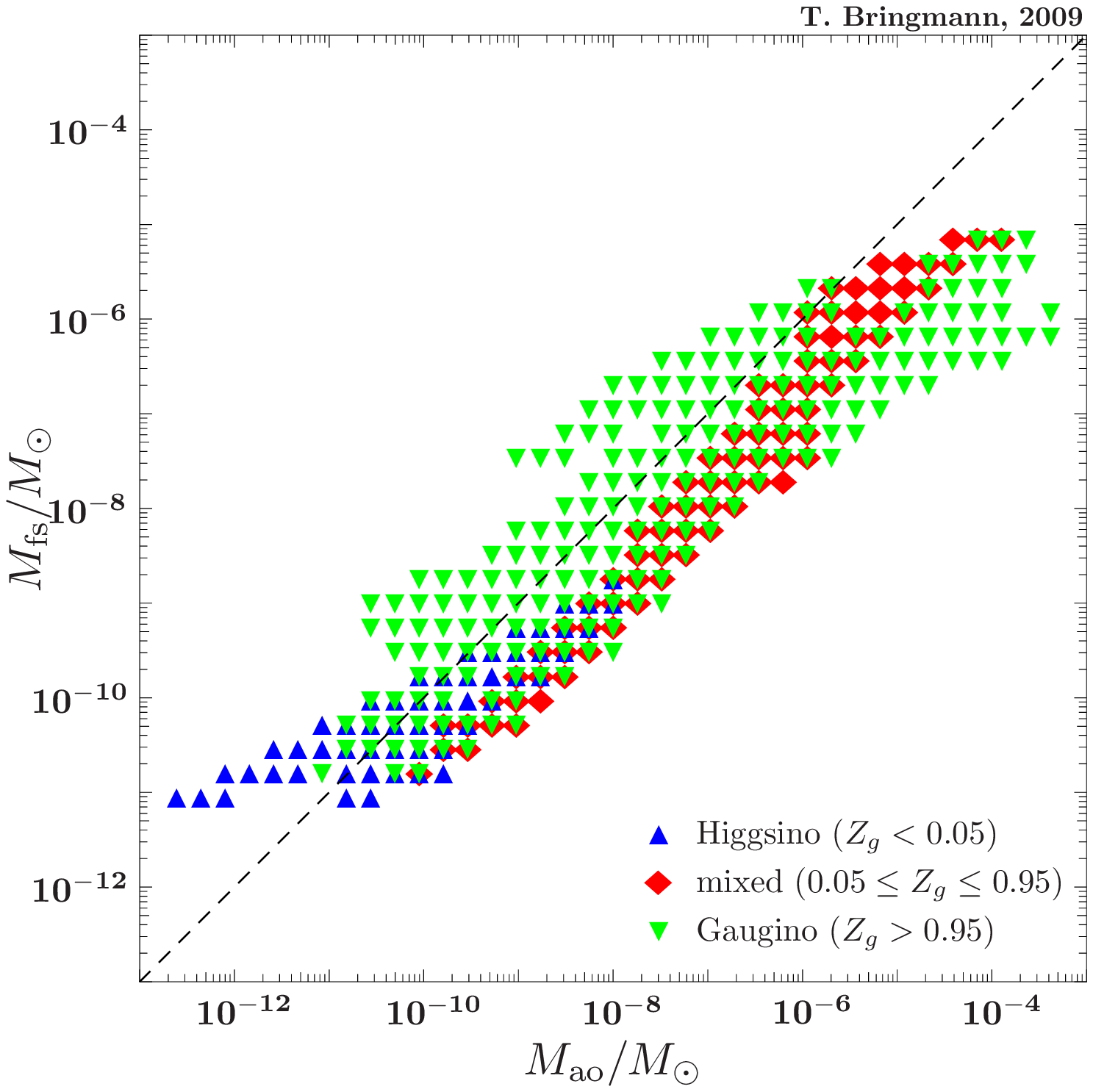}
   \includegraphics[width=0.486\columnwidth]{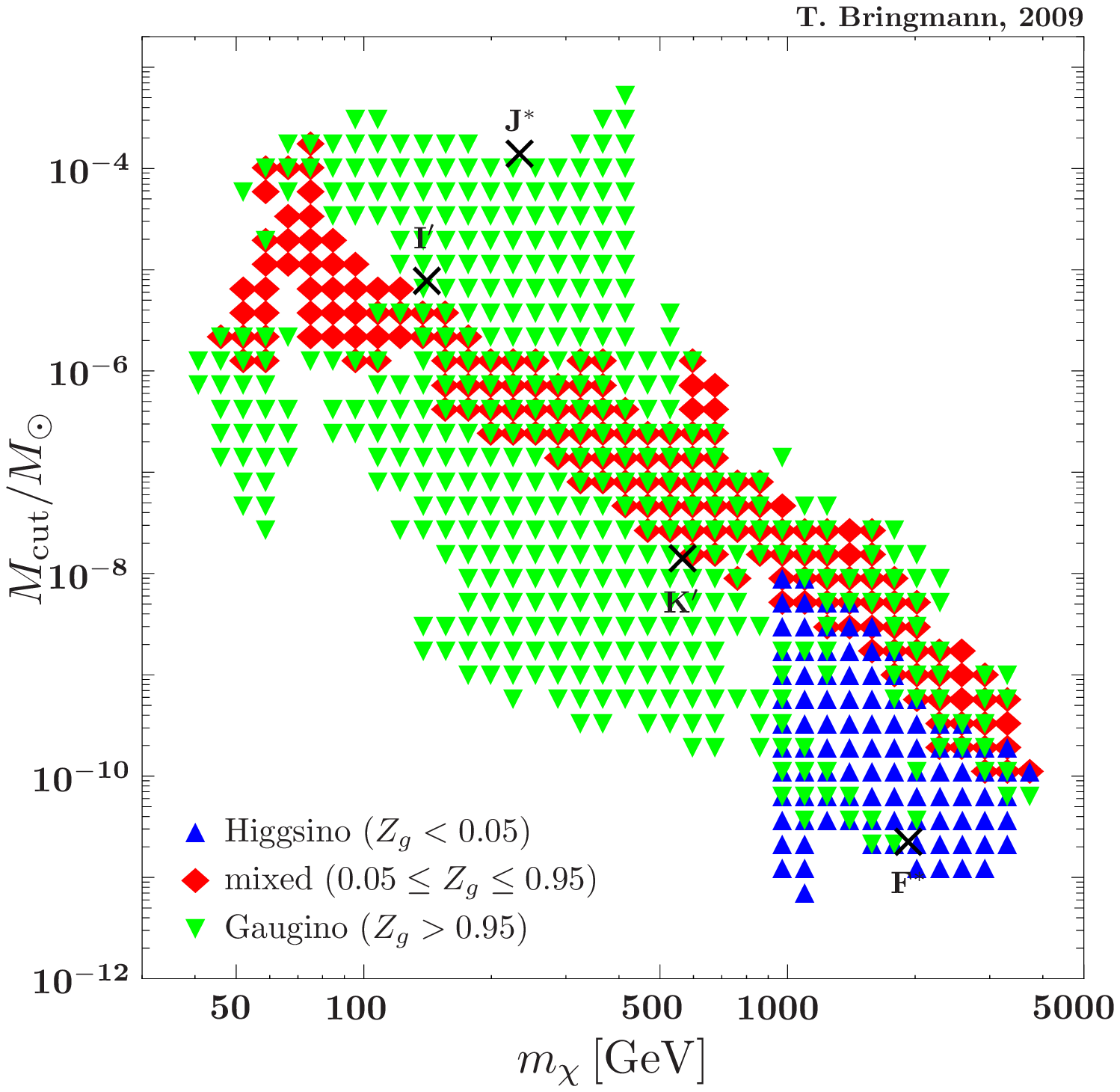}
\caption{The left panel shows the exponential cutoff scales associated to the main damping mechanisms of the matter power spectrum after kinetic decoupling, viz.~free streaming and the effect of acoustic oscillations, respectively; for models above (below) the dashed line, the former (latter) mechanism thus provides a stronger suppression of the power spectrum. In the right panel, the cutoff mass resulting from the dominating of these two independent effects is plotted against the neutralino mass, indicating the typical size of the smallest protohalos to be formed.
 \label{fig_Mcut}}
\end{figure}

The two cutoff scales are plotted against each other in the left panel of Fig.~\ref{fig_Mcut}. As noted before in \cite{Loeb:2005pm,Bertschinger:2006nq}, the effect of acoustic oscillations is, indeed, more important for the case of a Bino-like neutralino with $x_{\rm kd}\gtrsim 10^4$ (these models occupy the top part of the Figure, just to the right of the band containing mixed neutralinos). In general, however, this statement turns out to depend on the WIMP nature and there are cases where the free-streaming cutoff is almost 2 orders of magnitude above the one induced by acoustic oscillations. Since the two damping mechanisms are physically independent,
the actual cutoff $M_{\rm cut}$ in the power spectrum is thus, rather, given by $M_{\rm cut}=\max\left[M_{\rm fs},M_{\rm ao}\right]$; the possible range of $M_{\rm cut}$ is displayed in the right panel of Fig.~\ref{fig_Mcut} as a function of $m_\chi$. 

For very small values of $M_{\rm cut}$, corresponding to large $T_{\rm kd}$, one might wonder whether the QCD transition could leave an imprint on the power spectrum. In fact, if it is first order, the sound speed vanishes during the transition and density perturbations fall freely, potentially leading to the production of DM clumps with masses of $10^{-20}$ to $10^{-10}M_\odot$ \cite{Schmid:1998mx}. However, the corresponding enhancement factor in the CDM density fluctuations is only between 2 (from a lattice fit) and 20 (using the bag model) at scales of $\sim10^{-15}M_\odot$ and significantly smaller at larger scales; this has to be compared to the exponential suppression of power below $M_{\rm cut}$ due to the damping mechanisms discussed here. For the smallest cutoff scales shown in Fig.~\ref{fig_Mcut}, $M_{\rm cut}\lesssim10^{-10}M_\odot$, the actual cutoff mass might thus be slightly, but certainly not very much, smaller than indicated -- depending on the details of the QCD phase transition.

Following the paradigm of hierarchical structure formation, the smallest scales, and thus the scales closest to the cutoff, typically enter the non-linear regime first.
The smallest gravitationally bound objects to be formed in the universe are in that case also the first; protohalos  with a mass of around $M_{\rm cut}$. This behaviour has been confirmed numerically, where these protohalos could be followed until a redshift of $z\sim26$ \cite{Diemand:2005vz}. The range of expected minimal protohalo masses displayed in Fig.~\ref{fig_Mcut} is only slightly smaller than what was found earlier \cite{Profumo:2006bv} using an order-of-magnitude estimate for $T_{\rm kd}$ (based on \cite{Green:2005fa}) instead of the exact value as defined by the solution of Eq.~(\ref{eq:process}). For a given model, however, it turns out that the difference in the infered cutoff mass still is typically about a factor of 10, rather independent of $m_\chi$; adding to this the effect of identifying $M_{\rm ao}$ (like in \cite{Loeb:2005pm,Profumo:2006bv}) instead of $M_{\rm cut}=\max\left[M_{\rm fs},M_{\rm ao}\right]$ with the smallest protohalo mass, this difference can in some cases increase to a factor of almost 1000.

%%%%%%%%%%%%%%%%%%%%%%%%%%%%%%%%%%%%%%%%%%%%%%%%%%%%%%%%%%%%%%%%%%%%%%%%%%%%%%
\section{Discussion}
\label{sec:disc}
\subsection{Observational prospects}
\label{sec:obs}

Observational prospects depend crucially on whether the first protohalos survive until today or whether they are disrupted due to tidal interactions in merger processes or encounters with stars -- an issue that is still under debate. Several studies show that even if the protohalos lose some of their material on their way, most of the mass resides in a dense and compact core that remains intact \cite{Diemand:2005vz,clump_survival}. Other studies are less optimistic  \cite{Zhao:2005mb}. However, one should keep in mind that the very first objects probably form out of rare fluctuations with $M_{\rm rf}>M_{\rm cut}$ that enter the non-linear regime before fluctuations of size $M_{\rm cut}$; these DM clumps could develop denser cores before serious encounters, enhancing thus their survival probability. The importance of rare fluctuations for the formation of microhalos is also supported by numerical simulations \cite{Diemand:2005rd}.

A clumpy halo has important consequences for the indirect detection of DM, where one tries to discriminate exotic, DM-induced contributions to cosmic rays of various kinds from the standard astrophysical background. DM annihilating in dense clumps might, in particular, be visible in the form of gamma-ray point sources. In general, however, one can only expect to see some of the larger clumps \cite{clump_signals} with gamma-ray telescopes like Fermi/GLAST \cite{Baltz:2008wd}, while the very small clumps with masses close to $M_{\rm cut}$ are unlikely to be resolved \cite{Pieri:2005pg}. They do, however, contribute to the diffuse flux; indeed, it has been argued that models where Fermi is expected to see some of the larger clumps generally feature so many small clumps that their contribution to the diffuse flux would be in conflict with already existing observations \cite{Pieri:2007ir}. In any case does the so-called \emph{boost factor}, i.e.~the enhancement in the smooth component in a clumpy halo with respect to the flux expected for a homogenous DM distribution, depend on the small-scale cutoff of the clump-distribution -- which becomes even more important when one tries to fully take into account the self-similar nesting of halos (subhalos within subhalos). The actual functional dependence on $M_{\rm cut}$, however, strongly depends on the density profile of the smallest subhalos today as well as on their radial distribution in the Milky Way. Since the resolution of even the so far most ambitious numerical $N$-body simulations is still way above the expected cutoff scale, one has to extrapolate these results by many orders of magnitude, the implications of which are, understandably, still under debate. The most likely dependence of the boost factor on $M_{\rm cut}$ is logarithmic, i.e.~each decade in halo masses contributes about the same to the total boost. The boost factor for charged cosmic rays, conceptually a bit different from the one for gamma rays since charged particle propagation through the diffusive halo has to be taken into account, may be much less than for gamma rays \cite{Lavalle:1900wn}.

It has been proposed that the proper motion of the smallest subhalos could be detectable \cite{Koushiappas:2006qq}, but prospects for this idea do not look very promising even for the most favourable case of a rather large $M_{\rm cut}\sim10^{-2}M_\odot$; again, it is the EGRET diffuse background that places severe restrictions on this idea \cite{Ando:2008br}. While direct detection rates for DM would of course be greatly enhanced if a DM clump happens to pass through the earth, prospects are on average worse than for a smooth DM distribution \cite{Kamionkowski:2008vw}. Microhalos close to the cutoff scale, furthermore, feature a virial radius much larger than their Einstein radius, so gravitational lensing is commonly not considered an adequate method of assessing these smallest structures, either; recently, however, it has been suggested that a detailed study of multiple images of time-variable compact sources in strong lensing systems actually may have the potential to constrain the power spectrum in the lensing system even at scales close to typical values for the cutoff \cite{Moustakas:2009na}. 
 The maybe most promising way to directly probe the small-scale distribution of DM might be to use anisotropy probes such as the angular correlation in the extra-galactic gamma-ray background \cite{aniso} or the gamma-ray flux (one-point) probability function \cite{Lee:2008fm}. While galactic substructures do provide the dominant contribution to the first signal, they do so mostly for rather large clumps with masses $M\gtrsim10^4M_\odot$ \cite{Fornasa:2009qh}, which would imply that this method cannot be used to asses the small halos we are interested in here. The very recently proposed second method, however, claims to probe the subhalo distribution, at least in principle, down to the smallest scales.

To conclude this Section, let me mention that the role of microhalos in structure formation, and the implication of their presence for dark matter experiments, is also the subject of another review article in the present NJP focus issue on dark matter \cite{Koushiappas:2009du}.

%%%%%%%%%%%%%%%%%%%%%%%%%%%%%%%%%%%%%%%%%%%%%%%%%%%%%%%%%%%%%%%%%%%%%%%%%%%%%%
\subsection{Non-neutralino dark matter}
\label{sec:DM}

\begin{figure}[t]
   \begin{center}
   \includegraphics[width=0.7\columnwidth]{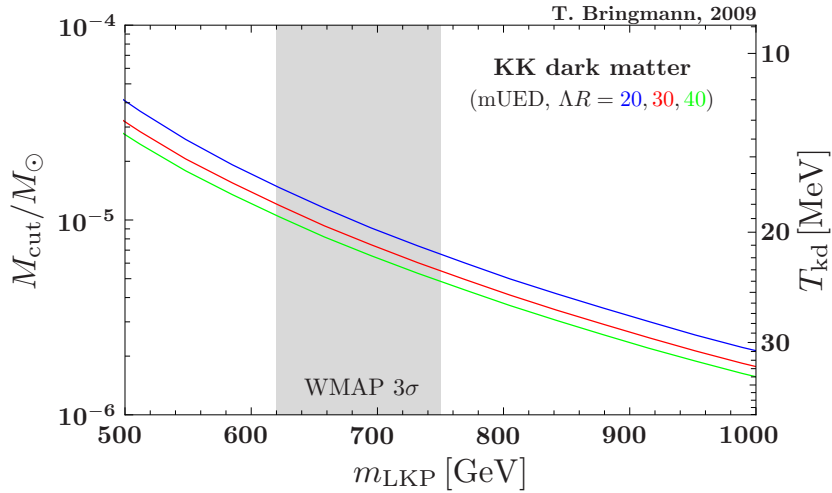}
    \end{center}
\caption{The kinetic decoupling temperature and the corresponding cutoff scale for the lightest Kaluza-Klein particle in universal extra dimensions. For high Higgs masses, $m_h\gtrsim150\,$GeV, the grey region corresponding to the relic density constraint given in Eq.(\ref{wmap})
shifts upwards, allowing an LKP mass up to around $1\,$TeV \cite{Kakizaki:2006dz}. 
 \label{fig_KK}}
\end{figure}

All expressions in Sections \ref{sec:dec}, \ref{sec:mcut} and \ref{app:coll} are equally applicable to \emph{any} WIMP candidate and the resulting decoupling temperatures and cutoff masses are expected to always fall into the range presented in the preceeding sections. In fact, in many cases the range will be much smaller simply because there are less free parameters to play with than in the case of supersymmetry. As an example, Fig.~\ref{fig_KK} shows the situation for Kaluza-Klein DM in universal extra dimensions \cite{ued}, where the mass splittings between the Kaluza-Klein particles arise due to a combination of electroweak and radiative contributions and are generally expected to be rather small. In the minimal setup \cite{Cheng:2002iz}, the whole spectrum of states can be specified by only two parameters; the compactification scale $R$ and the cutoff scale $\Lambda$ beyond which the four-dimensional effective theory ceases to be valid. For Kaluza-Klein dark matter, the cutoff in the matter power spectrum turns out to always be dominated by acoustic oscillations and, as can be seen in the Figure, falls into a very narrow range around $M_{\rm cut}=M_{\rm ao}\sim10^{-5}M_\odot$.

Another interesting class of DM candidates are particles that, unlike WIMPs, have never been in thermal equilibrium and for which the formalism presented here therefore does not apply (though it would be very interesting to extend it such as to cover even these cases). The maybe best studied example is the axion \cite{Sikivie:2006ni}. Initially misaligned, it starts to oscillate coherently around $T\sim1\,$GeV and from then on evolves as $\rho_a\propto a^{-3}$ just like ordinary CDM. Inhomogeneities on scales larger than the Hubble horizon at the temperature of realignment later evolve into axion mini-clusters with a typical mass around 
$10^{-12}M_\odot$ \cite{axioncut}. One should, however, keep in mind that for particles that decouple before the QCD transition -- like axions, primordial black holes with $M_{\rm PBH}\ll M_\odot$ or ultra-cold WIMPs \cite{Gelmini:2008sh} -- small-scale fluctuations may be strongly amplified if the transition is first order, producing DM clumps with masses of $10^{-20}$ to $10^{-10}M_\odot$ \cite{Schmid:1998mx}.

Finally, if DM consists of superWIMPs that result from the late decay of thermally produced WIMPs, the actual cutoff in the power-spectrum is not the one from the WIMP decoupling but the one that is imposed from the kinematics of the decay (through the mass difference between decaying particle and DM particle). In fact, such models have been proposed to address a certain tension that is sometimes claimed at ``small'' scales (in this case Mpc instead of the pc scales that correspond to $M_{\rm cut}\sim10^{-5}M_\odot$) between observations and numerical $N$-body simulations \cite{latedecay}. However, this idea works only partially \cite{Borzumati_et_al.}; what is more, the evidence for small-scale ``problems'' of standard $\Lambda$CDM cosmology may soon well disappear completely, with more detailed observations and $N$-body simulations starting to converge \cite{Simon_Geha}. Nevertheless, late-decaying DM is an interesting possibility that does not have to be related to this particular idea; in contrast to the typical $M_{\rm cut}$ for WIMPs, a large cutoff in the power spectrum might even be possible to probe by future micro-lensing missions.

%%%%%%%%%%%%%%%%%%%%%%%%%%%%%%%%%%%%%%%%%%%%%%%%%%%%%%%%%%%%%%%%%%%%%%%%%%%%%%
\section{Conclusions}
\label{sec:con}

The kinetic decoupling process of WIMPs from the thermal bath can be followed in great detail by solving the full Boltzmann equation in this regime. Extending the formalism presented in \cite{BH}, by allowing for non-relativistic scattering partners and taking into account the full time-dependence of the effective number of degrees of freedom, a highly precise determination of the decoupling temperature becomes possible that in turn can be translated into a small-scale cutoff in the spectrum of matter density fluctuations.

An extensive scan over the parameter space for supersymmetric neutralino DM reveals a slightly smaller range of cutoff masses, $10^{-11}M_\odot$ to a few times $10^{-4}M_\odot$, but basically confirms the only existing corresponding scan so far \cite{Profumo:2006bv}, which is based on an order-of-magnitude estimate for the decoupling temperature (given in \cite{Green:2005fa}). The resulting difference in $M_{\rm cut}$ for individual WIMP models, however, can be sizable; typically of the order of $10$, models with a difference of almost $10^3$ were found. Another important result of the scan presented here is that whether free streaming or acoustic oscillations are more effective in the suppression of power on small scales depends on the DM particle nature (in slight disagreement to the claim of \cite{Bertschinger:2006nq} who presented a very detailed study of the evolution of density contrasts through and after kinetic decoupling, albeit based only on one particular DM candidate).

The range of decoupling temperatures and cutoff masses presented here is indicative for the whole class of WIMP DM candidates, though many models -- such as Kaluza-Klein DM -- will exhibit a much smaller range. For non-WIMP candidates, the mass of the smallest clumps can differ significantly from the range derived here; it would be interesting to develop tools that allow an as precise determination of the cutoff scale for these cases as for the case of WIMPs. As for detectional prospects of the smallest DM clumps, many interesting ideas have been put forward. Though challenging, it is an exciting possibility that one may be able to measure the DM distribution on such scales in the future. In order to really address the connection to the microphysics of the DM particles, however, one still needs a better understanding of how the first protohalos evolve and, given their initial distribution, what they are expected to look like today.

The routines for calculating the kinetic decoupling temperature and the associated cutoff scale have been implemented in \ds\ \cite{ds} and will be available with the next release \cite{dsupdate}.

%\section*{Acknowledgments}
%\ack

%%%%%%%%%%%%%%%%%%%%%%%%%%%%%%%%%%%%%%%%%%%%%%%%%%%%%%%%%%%%%%%%%%%%%%%%%%%%%%
\appendix
%%%%%%%%%%%%%%%%%%%%%%%%%%%%%%%%%%%%%%%%%%%%%%%%%%%%%%%%%%%%%%%%%%%%%%%%%%%%%%

%%%%%%%%%%%%%%%%%%%%%%%%%%%%%%%%%%%%%%%%%%%%%%%%%%%%%%%%%%%%%%%%%%%%%%%%%%%%%%
\section{The collision term in the Boltzmann equation}
\label{app:coll}

In this Appendix, we derive the collision term for scattering processes between non-relativistic WIMPs $\chi$ and considerably less massive (usually SM) particles that are in thermal equilibrium with the plasma in the early universe. For the former, we use $p^\mu=(E,\mathbf{p})$ to denote ingoing momenta, while for the latter we use $k^\mu=(\omega,\mathbf{k})$; the corresponding outgoing quantities are marked with a tilde. The treatment here follows closely that of Ref.~\cite{BH}, but slightly extends it in taking into account the possibility of non-relativistic scattering partners.

The collision term, i.e.~the right-hand side of the Boltzmann equation, reads
\be
  \label{Cfull}
  \fl C=\int\frac{d^3k}{(2\pi)^32\omega}\int\frac{d^3\tilde k}{(2\pi)^32\tilde \omega}\int\frac{d^3\tilde p}{(2\pi)^32\tilde E}(2\pi)^4\delta^{(4)}(\tilde p+\tilde k-p-k)\overline{\left|\mathcal{M}\right|}^2J\,,
\ee
where 
\bea
 \label{Jdef}
 J&\equiv& g_\mathrm{SM}\left[\left(1\mp g^\pm(\omega)\right)\, g^\pm(\tilde\omega)f(\mathbf{\tilde p})-\left(1\mp g^\pm(\tilde\omega)\right)\, g^\pm(\omega)f(\mathbf{p})\right]\,,\\
  g^\pm(\omega) &=& \left(e^{\omega/T}\pm1\right)^{-1}\,.
\eea
Here, $\overline{\left|\mathcal{M}\right|}^2$ is the scattering amplitude squared, summed over final and averaged over initial spin states, and
$g_\mathrm{SM}$ is the number of internal degrees of freedom of the scattering partner (the upper signs apply for fermions, the lower for bosons); as for the following expressions, a summation over all SM particles in thermal equilibrium is always understood.
No assumptions about the $\chi$ distribution function $f(\mathbf{p})$ are necessary;  as long as the WIMPs are much less abundant than their scattering partners, however, Pauli suppression factors for $f$ can safely be neglected -- as has been done in Eq.(\ref{Jdef}).

After chemical freezeout, one typically has $\omega\sim T\ll m_\chi$. For kinematical reasons, the average momentum transferred during the scattering events is thus small, so \kref{Cfull} can be expanded as 
\bea
   \label{Cexpansion}
   C(E)&=& \sum_{j=0}^{\infty}C^j\,,\\
   C^j&\equiv&\int\frac{d^3k}{(2\pi)^32\omega}\int\frac{d^3\tilde k}{(2\pi)^32\tilde \omega}\int\frac{d^3\tilde p}{(2\pi)^32\tilde E}\nonumber\\
     &&\times(2\pi)^4\delta(\tilde E+\tilde\omega-E-\omega)\overline{\left|\mathcal{M}\right|}^2J\left[\frac{1}{j!}D_\mathbf{q}^j(\mathbf{\tilde p})\delta^{(3)}(\mathbf{\tilde p}-\mathbf{p})\right]\,,
\eea
where
\be
  D_\mathbf{q}(\mathbf{\tilde p}) \equiv \mathbf{q}\cdot\nabla_\mathbf{\tilde p}\equiv(\mathbf{\tilde k}-\mathbf{k})\cdot\nabla_\mathbf{\tilde p}\,,
\ee
 and the derivatives of the delta function are as usual defined through an integration by parts.
Following similar steps as detailed in Ref.~\cite{BH}, but paying special attention to the difference between $\omega$ and $k=\left|\mathbf{k}\right|$, one can now proceed to calculate the expansion coefficients $C^j$. To lowest non-vanishing order in $\mathbf{p}^2/E^2$ and $\omega/m_\chi$, one then arrives at the following expression for the collision integral:
\be
  C= C^1+C^2 = c(T) m_\chi^2 \Big[m_\chi T\Delta_\mathbf{p} +\mathbf{p}\cdot\nabla_\mathbf{p}+3\Big]\,f(\mathbf{p})\,,
\ee
where
\be
  \label{cTdef}
  c(T) =  \sum_i\frac{g_\mathrm{SM}}{6(2\pi)^3m_\chi^4T} \int dk\,k^5 \omega^{-1}\,g^\pm\left(1\mp g^\pm\right)\mathop{\hspace{-11ex}\overline{\left|\mathcal{M}\right|}^2_{t=0}}_{\hspace{4ex}s=m_\chi^2+2m_\chi\omega+m_\ell^2}\,.
\ee
For clarity, the sum over all SM scattering partners $i$ has here been made explicit.

As an interesting aside, the above integral can be solved analytically in the relativistic limit, $m_\ell\rightarrow0$, if the amplitude squared scales like a power of $\omega$ \cite{BH}. For WIMP scattering below $s$-channel resonances, $m_\ell\ll\omega\ll\omega_{\rm res}$, e.g., the scattering amplitude is always given by $\overline{\left|\mathcal{M}\right|}^2=\overline{\left|\mathcal{M}\right|}^2_0\,(\omega/m_\chi)^2$, see \ref{app:nfscatter}. In this particular case, the integral in Eq.~(\ref{cTdef}) becomes:
\be
  \int dk\,k^5 \omega^{-1}\,g^\pm\left(1\mp g^\pm\right)\overline{\left|\mathcal{M}\right|}^2
\stackrel{m_\ell\ll\omega\ll\omega_{\rm res}}{\longrightarrow}
720\,\zeta(7)\,\overline{\left|\mathcal{M}\right|}^2_0\,m_\chi^{-2}\,T^7
\ee
for bosonic scattering partners; for fermions, the above expression has to be multiplied by $31/32$.

%%%%%%%%%%%%%%%%%%%%%%%%%%%%%%%%%%%%%%%%%%%%%%%%%%%%%%%%%%%%%%%%%%%%%%%%%%%%%%
\section{Neutralino-fermion scattering amplitude}
\label{app:nfscatter}

The scattering between neutralinos and fermions is mediated by sfermion exchange in the $s$- and $u$-channel, and $Z$- and Higgs-boson exchange in the $t$-channel. Denoting with $p$ ($\tilde p$) the ingoing (outgoing) momentum of the lightest neutralino $\chi_1$, and with $k$ ($\tilde k$) the ingoing (outgoing) momentum of the fermionic scattering partner $\ell$, the total scattering amplitude is given by

\be
  \mathcal{M} = \mathcal{M}_s + \mathcal{M}_u + \mathcal{M}_t\,,
\ee
with
\bea
 \fl \mathcal{M}_s = i\sum_{i=1,2} \Delta_s(\tilde\ell_i)\,
      u^T_pC^{-1}\left\{g^{R^*}_{\tilde\ell_i\ell1}P_L+g^{L^*}_{\tilde\ell_i\ell1}P_R\right\} u_k\,
       \bar{u}_{\tilde k}\left\{g^L_{\tilde\ell_i\ell1}P_L+g^R_{\tilde\ell_i\ell1}P_R\right\} C\bar{u}^T_{\tilde p}\\
\fl \mathcal{M}_u = -i\sum_{i=1,2} \Delta_u(\tilde\ell_i)\,
      \bar{u}_{\tilde k}\left\{g^L_{\tilde\ell_i\ell1}P_L+g^R_{\tilde\ell_i\ell1}P_R\right\} u_p\,
       \bar{u}_{\tilde p}\left\{g^{R^*}_{\tilde\ell_i\ell1}P_L+g^{L^*}_{\tilde\ell_i\ell1}P_R\right\} u_{k}\\
\fl \mathcal{M}_t = i\Delta_t(Z)\,
      \bar{u}_{\tilde p}g^L_{Z11}\gamma_5\gamma_\mu u_{p}\,
      \bar{u}_{\tilde k}\gamma^\mu\left\{g^L_{Z\ell\ell}P_L+g^R_{Z\ell\ell}P_R\right\} u_{k}\nonumber\\
  -i\sum_{m=1,2}\Delta_t(H_s)\,
      \bar{u}_{\tilde p}\left\{\Re\left[g^L_{H_m11}\right]-i\gamma_5\Im\left[g^L_{H_m11}\right]\right\} u_{p}\,
\bar{u}_{\tilde k} g^L_{H_m\ell\ell} u_{k}\nonumber\\    
  +i\Delta_t(H_3)\,
      \bar{u}_{\tilde p}\left\{\Re\left[g^L_{H_311}\right]-i\gamma_5\Im\left[g^L_{H_311}\right]\right\} u_{p}\,
\bar{u}_{\tilde k} g^L_{H_3\ell\ell} \gamma_5 u_{k}\,,
\eea     
and
\be
  \Delta_{(s,t,u)}(A) \equiv -\left[(s,t,u)-m_{A}^2+im_\chi\Gamma_{A}\right]^{-1}
\ee
for any particle or sparticle $A$ with mass $m_A$ and width $\Gamma_A$.
In the above expressions, $H_1$ and $H_2$ denote the scalar Higgs-bosons and $H_3$ its pseudoscalar version, $C$ the charge conjugation matrix and a superscript $T$ the transpose of a matrix. For the coupling constants, the same conventions as in \ds\ \cite{ds} are adopted (for an explicit representation of these couplings, see e.g.~\cite{Edsjo:1997hp}; for a nice summary of Feynman rules for Majorana fermions, see \cite{Haber:1984rc}). In the case of neutrino scattering, there is only one sneutrino to be exchanged; one thus has to drop the corresponding sums over sfermion mass eigenstates ($i,j$) and to replace $\tilde\ell_{i,j}\rightarrow\tilde\nu$ in the above and the following expressions (in this case, one may of course also safely neglect the exchange of Higgs bosons).

The typical CMS fermion energy is much smaller than the neutralino mass, i.e.~$\omega\ll m_\chi$. As explained in \ref{app:coll}, we are therefore only interested in small momentum transfer,
\bea
  t&\rightarrow&0\,,\\
  s&\rightarrow& m_\chi^2+2m_\chi\omega+m_\ell^2\,.
\eea
In this limit, averaging over initial and summing over final state spins gives:
\be
  \label{m2full}
  \fl \overline{\left|\mathcal{M}\right|}^2=
      \overline{\left|\mathcal{M}_s\right|}^2+ \overline{\left|\mathcal{M}_t\right|}^2+ \overline{\left|\mathcal{M}_u\right|}^2+2\Re\left[\overline{\mathcal{M}_s\mathcal{M}_t^*}-\overline{\mathcal{M}_s\mathcal{M}_u^*}-\overline{\mathcal{M}_t\mathcal{M}_u^*}\right]\,,
\ee

\noindent
where
\bea
\fl \overline{\left|\mathcal{M}_s\right|}^2 = 
m_\chi^2\sum_{i,j=1,2} \Delta_s(\tilde\ell_i)\Delta^*_s(\tilde\ell_j)
  \left|\omega\left\{g^{L^*}_{\tilde\ell_i\ell1}g^{L}_{\tilde\ell_j\ell1}+g^{R^*}_{\tilde\ell_i\ell1}g^{R}_{\tilde\ell_j\ell1}\right\}
  -m_\ell \left\{g^{L^*}_{\tilde\ell_i\ell1}g^{R}_{\tilde\ell_j\ell1}+g^{R^*}_{\tilde\ell_i\ell1}g^{L}_{\tilde\ell_j\ell1}\right\}\right|^2\nonumber\\
\\
\fl \overline{\left|\mathcal{M}_u\right|}^2 =
\overline{\left|\mathcal{M}_s\right|}^2(\omega\rightarrow-\omega)\\
\fl \overline{\left|\mathcal{M}_t\right|}^2=
4m_\chi^2\left|\Delta_t(Z)\right|^2\left(g^{L}_{Z11}\right)^2\Big[
\left\{g^{L^2}_{Z\ell\ell}+g^{R^2}_{Z\ell\ell}\right\}(2\omega^2+m_\ell^2)
-6g^{L}_{Z\ell\ell}g^{R}_{Z\ell\ell}m_\ell^2
\Big]\nonumber\\
\fl\qquad\qquad+16m_\ell^2m_\chi^2\sum_{l,m=1,2}\Delta_t(H_l)\Delta^*_t(H_m)g^L_{H_l\ell\ell}g^L_{H_m\ell\ell}\Re\left[g^L_{H_l11}\right]\Re\left[g^L_{H_m11}\right]\\
\fl \overline{\mathcal{M}_s\mathcal{M}_u^*} =
m_\chi^2\sum_{i,j=1,2} \Delta_s(\tilde\ell_i)\Delta^*_u(\tilde\ell_j)\Bigg[
2\Im\left[g^{L^*}_{\tilde\ell_i\ell1}g^{R}_{\tilde\ell_i\ell1}\right]\Im\left[g^{L^*}_{\tilde\ell_j\ell1}g^{R}_{\tilde\ell_j\ell1}\right](\omega^2-m_\ell^2)\nonumber\\
\fl\qquad\qquad\qquad -\frac{m_\ell^2}{2}\left\{\left|g^{L}_{\tilde\ell_i\ell1}\right|^2+\left|g^{R}_{\tilde\ell_i\ell1}\right|^2-\frac{2\omega}{m_\ell}\Re\left[g^{L^*}_{\tilde\ell_i\ell1}g^{R}_{\tilde\ell_i\ell1}\right]\right\}\nonumber\\
\fl\qquad\qquad\qquad\qquad
\times\left\{\left|g^{L}_{\tilde\ell_j\ell1}\right|^2+\left|g^{R}_{\tilde\ell_j\ell1}\right|^2+\frac{2\omega}{m_\ell}\Re\left[g^{L^*}_{\tilde\ell_j\ell1}g^{R}_{\tilde\ell_j\ell1}\right]\right\}
\Bigg]\\
\fl\overline{\mathcal{M}_s\mathcal{M}_t^*} =
m_\chi^2 \sum_{i=1,2} \Delta^*_t(Z)\Delta_s(\tilde\ell_i)
g^{L}_{Z11}\Bigg[3g^{R}_{Z\ell\ell}\left|g^{L^*}_{\tilde\ell_i\ell1}\omega-g^{R}_{\tilde\ell_i\ell1}m_\ell\right|^2
-3g^{L}_{Z\ell\ell}\left|g^{L^*}_{\tilde\ell_i\ell1}m_\ell-g^{R}_{\tilde\ell_i\ell1}\omega\right|^2\nonumber\\
\fl\qquad\qquad\qquad
+\left\{g^{L}_{Z\ell\ell}\left|g^{R}_{\tilde\ell_i\ell1}\right|^2-g^{R}_{Z\ell\ell}\left|g^{L}_{\tilde\ell_i\ell1}\right|^2\right\}\left(\omega^2-m_\ell^2\right)
\Bigg]\nonumber\\
\fl\qquad\qquad+2m_\chi^2\sum_{i=1,2}\sum_{m=1,2} \Delta_s(\tilde\ell_i)\Delta^*_t(H_m)g^L_{H_m\ell\ell}\Re\left[g^L_{H_m11}\right]\Bigg[\left\{\left|g^{L}_{\tilde\ell_i\ell1}\right|^2+\left|g^{R}_{\tilde\ell_i\ell1}\right|^2\right\}m_\ell \omega\nonumber\\
\fl\qquad\qquad\qquad -2\Re\left[g^{L^*}_{\tilde\ell_i\ell1}g^{R}_{\tilde\ell_i\ell1}\right]m_\ell^2
\Bigg]\\
\fl\overline{\mathcal{M}_t\mathcal{M}_u^*} =
-\overline{\mathcal{M}_t\mathcal{M}_s^*}(\omega\rightarrow-\omega)
\eea
The above expressions extend the formulae given in  \cite{Chen:2001jz} for neutralino-neutrino scattering and correctly reproduce the expressions obtained in \cite{Griest:1988ma} for neutralino annihilation (after the replacements $t\rightarrow u, u\rightarrow s, s\rightarrow u$).

For relativistic fermions, the above expression further simplifies considerably. Below the resonance(s)
\be
  \omega_{\rm res, i}\equiv\frac{m_{\tilde\ell_i}^2-m_\chi^2}{2m_\chi}\,,
\ee
the scattering amplitude becomes
\bea
   \label{m2simp}
   \fl \overline{\left|\mathcal{M}\right|}^2_{m_\ell\ll\omega\ll\omega_{\rm res}} \approx 2 \left(\frac{\omega}{m_\chi}\right)^2
\Bigg\{
4\left(\frac{m_\chi}{m_Z}\right)^4g^{L^2}_{Z11}
\left[g^{L^2}_{Z\ell\ell}+g^{R^2}_{Z\ell\ell}\right]
\\
\qquad\quad+\sum_{i,j=1,2}\frac{\left|g^{L}_{\tilde\ell_i\ell1}\right|^2\left|g^{L}_{\tilde\ell_j\ell1}\right|^2+\left|g^{R}_{\tilde\ell_i\ell1}\right|^2\left|g^{R}_{\tilde\ell_j\ell1}\right|^2 }
{\left(1-m_{\tilde\ell_i}^2/m_\chi^2\right)\left(1-m_{\tilde\ell_j}^2/m_\chi^2\right)}\Bigg\}
\,.\nonumber
\eea
Above the resonance, $\omega_{\rm res}\ll\omega\ll m_\chi$, the $s$- and $u$-channel amplitudes are roughly constant, while the $t$-channel amplitude continues to scale like $\mathcal{M}_t\propto\omega$.
In the limit of a pure Bino (i.e.~$\left|g^L_{\tilde\ell_R\ell1}\right|=\sqrt{2}g_YY_s$, $\left|g^R_{\tilde\ell_L\ell1}\right|=\sqrt{2}g_YY_d$,$g^L_{Z11}=0$), and degenerate sfermions,  Eq.(\ref{m2simp}) becomes the expression earlier found in \cite{Hofmann:2001bi}:
\be
  \label{m2bino}
  \overline{\left|\mathcal{M}\right|}^{2~~{\rm(Bino)}}_{m_\ell\ll\omega\ll\omega_{\rm res}} \approx 8g_Y^4\left(Y_d^4+Y_s^4\right)\omega^2\left(\frac{m_\chi}{m_\chi^2-m_{\tilde\ell}^2}\right)^2\,.
\ee
Let me stress that the expressions (\ref{m2simp}) and (\ref{m2bino}) are just given here for convenience; for all numerical calculations, the full expression (\ref{m2full}) was used.

\end{document}